\documentclass[%
aps,
preprint,
showpacs,
preprintnumbers,
 amsmath,
 amssymb,
prl,
]{revtex4-1}

\usepackage{graphicx}
\usepackage{dcolumn}
\usepackage{bm}

\begin{document}

\title{Cascaded acceleration of proton beams in ultrashort laser-irradiated microtubes}

\author{H. C. Wang$^{1,2}$}
\author{S. M. Weng$^{1,2}$}\email{wengsuming@gmail.com}%
\author{M. Murakami$^{3}$}
\author{Z. M. Sheng$^{1,2,4}$}\email{z.sheng@strath.ac.uk}%
\author{M. Chen$^{1,2}$}
\author{Q. Zhao$^{1,2}$}
\author{J. Zhang$^{1,2}$}

\affiliation{$^{1}$ Key Laboratory for Laser Plasmas (MoE), School of Physics and Astronomy,
Shanghai Jiao Tong University, Shanghai 200240, China}%
\affiliation{$^{2}$ Collaborative Innovation Center of IFSA, Shanghai Jiao Tong University,
Shanghai 200240, China}%
\affiliation{$^{3}$ Institute of Laser Engineering, Osaka University, Osaka 565-0871, Japan}
\affiliation{$^{4}$ SUPA, Department of Physics, University of Strathclyde, Glasgow G4 0NG, UK}

\date{\today}
\begin{abstract}
A cascaded ion acceleration scheme is proposed by use of ultrashort laser-irradiated microtubes.
When the electrons of a microtube are blown away by intense laser pulses, strong charge-separation electric fields are formed in the microtube both along the axial and along the radial directions.
By controlling the time delay between the laser pulses and a pre-accelerated proton beam injected along the microtube axis, we demonstrate that this proton beam can be further accelerated by the transient axial electric field in the laser-irradiated microtube.
Moreover, the collimation of the injected proton beam can be enhanced by the inward radial electric field.
Numerical simulations show that this cascaded ion acceleration scheme works efficiently even at non-relativistic laser intensities, and it can be applied to injected proton beams in the energy range from 1 to 100 MeV.
Therefore, it is particularly suitable for cascading acceleration of protons to higher energy.
\end{abstract}

\pacs{52.38.Kd, 78.67.Ch, 41.75.Jv, 52.65.Rr}

\maketitle

\section{Introduction}

Plasma-based ion acceleration driven by intense laser pulses attracts growing attention over the last decades due to their wide range of applications in fundamental science, medicine, and industry \cite{DaidoRPP2012,MacchiRMP2013}.
Principally, laser-driven ion acceleration results from the charge separation in a double layer produced in the laser-target interaction \cite{HoraLPB1983}. Such a charge separation can lead to an accelerating field as large as a few hundred of GV/m, which is three orders of magnitude higher than the maximum accelerating field achieved in the conventional accelerators driven by radio frequency (RF) fields \cite{WilksPOP2001}.
This dramatic enhancement of the accelerating field brings out the advantages of more compact size, and higher particle density in the laser-driven ion acceleration.
These advantages are of great benefit to many applications, such as radiography and radiotherapy \cite{BorghesiPOP2002,Medphys2008,Loeffler}, high energy density physics \cite{Patel,Tahir,Xu2012} and so on.
In particular, a thick plasma block can be accelerated as a whole by an intense laser pulse if its duration is long enough \cite{HoraPOP2007,HoraPLA2013,HoraLPB2014,Weng2014PoP,WengSR}.
The generated plasma block, having a ultrahigh energy fluence, is of great benefit to novel ignition schemes of inertial confinement fusion, such as fast ignition and block ignition \cite{RothPRL2001,HoraBook}.

Stimulated by these prospective applications, enormous progresses have been achieved in both the experimental and theoretical study of laser-driven ion acceleration.
For instance, the proton cut-off energy of 85 MeV with high particle numbers has been demonstrated in the recent experiment \cite{WagnerPRL2016} via the target-normal sheath acceleration (TNSA), which is the most studied mechanism for laser-driven ion acceleration.
In the double-target scheme proposed recently, it was found that the charge separation field in the main target can be greatly enhanced by introducing a relatively low-density pre-target \cite{XuAIP2016,XuPOP2017,LiPOP2017}. As a result, the energy, quality and ion number of the accelerated ion beam can be all enhanced.
However, laser-driven ion acceleration are usually realized via the charge-separation field, which only exists in an extremely narrow region of a thickness of a few microns.
With such a short accelerating distance, it is very challenging to get higher ion energy.
To enhance the ion energy, the radiation pressure acceleration (RPA) has been proposed \cite{RPApapers1,RPApapers2,BulanovPOP2010}, in which the ions of an ultrathin foil can be accelerated continuously as a sail since they move together with the charge-separation field.
Theoretically, there is no limitation of the achievable ion energy and a very high conversion efficiency can be realized in the relativistic limit.
However, there is a large gap between the theoretical predictions and experimental results for RPA due to its extremely stringent requirements in laser and target conditions \cite{BinPRL2015,KimPoP2016}.
For instance, it is still challenging to generate a high-power circularly-polarized laser pulse \cite{WengOptica}. The latter is critically required for suppressing the plasma heating in RPA.
In addition, the hole-boring effect and the transverse instabilities will distort the ultrathin target and hence terminate the acceleration \cite{RPApapers1,RPApapers2}.

\begin{figure}[htbp]
\includegraphics[width=0.75\textwidth]{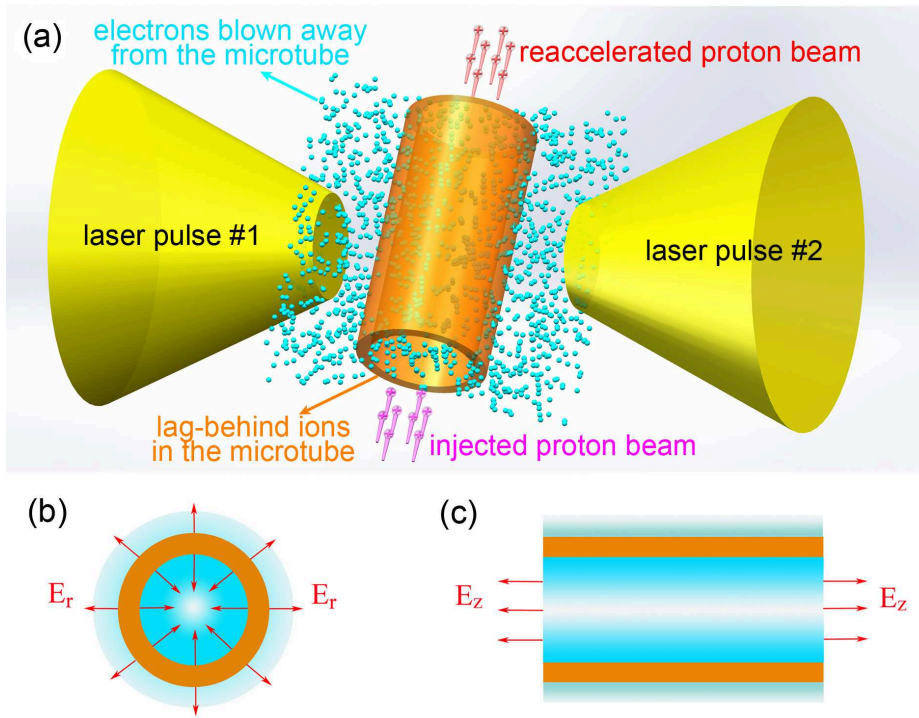}
\caption{(color online).
Schematic of a cascaded ion acceleration scheme. (a) An injected proton beam is further accelerated after it passes through a mictrotube that irradiated by two laser pulses along the radial direction.
The expanding electrons and lag-behind ions of the mictrotube combine to generate (b) a radial electric field $E_r$ and (c) an axial field $E_z$.} \label{figSchematic}
\end{figure}

On the other hand, one notices that the conventional accelerators have achieved remarkable successes regardless of their relatively weaker accelerating fields. About 30,000 conventional accelerators have been built worldwide to supply high-quality high-energy particle beams for the applications from fundamental researches to medicinal and industrial fields \cite{Feder}.
The secret to conventional accelerators' success is that the particles can pass through the accelerating field multiple times, so the output energy is not limited by the strength of the accelerating field \cite{Humphries}.
Inspired by this, a few cascaded laser-driven ion acceleration schemes have been proposed \cite{Pfotenhauer,Wang,Pei}. It was numerically demonstrated that via these cascaded schemes the energies of the injected proton beams can be doubled approximately, while rare attention is paid to the beam quality such as the collimation \cite{Kawata,Xu2014}.

We propose a cascaded ion acceleration scheme using laser-irradiated microtubes, by which both the energy and the quality of an injected proton beam can be enhanced.
Irradiated by short intense laser pulses, a microtube will be ionized instantaneously and its electrons will be quickly blown away as shown in Fig. \ref{figSchematic}(a).
Due to the expansion of electrons a radial charge-separation electric field will be generated, which is inward inside the microtube as illustrated in Fig. \ref{figSchematic}(b). So it can play a unique role in the focusing of ion beams \cite{ToncianScience2006}.
While the lag-behind ions will form a positively charged hollow cylinder, which brings out a strong axial electric field outward along the axis as displayed in Fig. \ref{figSchematic}(c).
It was proved that this axial electric field can be utilized to generate a well-collimated quasi-monoenergetic proton beam, whose divergence angle is less than 1 degree and energy spread $\Delta E/E \sim 10\%$\cite{MurakamiAPL2013}.
However, the energy of generated protons is only about a few MeV \cite{MurakamiAPL2013}, which greatly limits its applications.
By controlling the arrival time of an injected pre-accelerated proton beam in Fig. \ref{figSchematic}(a), we will show that this proton beam can be efficiently gained energy after it passes through the laser-irradiated microtube.
We demonstrate that this cascaded ion acceleration scheme works well with the injected proton beams of energies up to $100$ MeV.
More importantly, both the divergence and energy spread of the injected proton beam can be maintained at a low level, or even be suppressed.
Therefore, this scheme is particularly suitable to be cascaded to accelerate protons repeatedly. 

\section{Theoretical and simulation results}
To verify this cascaded ion acceleration scheme, we have performed 3D3V particle-in-cell (PIC) simulations using the code Osiris \cite{Fonseca}.
In simulations, two linearly polarized laser pulses irradiate a microtube along the $x$-axis from two sides.
Each pulse has a peak intensity $I\simeq 5.35 \times10^{19}$ W/cm$^2$ (the normalized vector potential $a\equiv|e\textbf{E}/\omega m_e c|=5$ and the laser wavelength $\lambda=0.8 \mu m$), a Gaussian transverse profile with a spot radius $3\lambda$, and a trapezoidal ($2T_0$ rise + $10T_0$ plateau + $2T_0$ fall) temporal profile, where $T_0=2\pi/\omega$ is the wave period.
The microtube has a length $2L=2\lambda$ along the $z$-axis, an inner radius $R=0.2\lambda$, a thickness $D=0.05\lambda$, and a uniform electron density $n_e=100n_c$, where $n_c \simeq 1.74\times 10^{21}$ cm$^{-3}$ is the critical density.
Such kind of a microtube could be fabricated via the deposition of gold atoms into the monolayer of carbon atoms of a carbon nanotube \cite{Kim2006,Smith1998Nature,Zhao1997Carbon}.
At the laser intensity $I\simeq 5.35 \times10^{19}$ W/cm$^2$, the gold atoms will be photoionized to a state of about $Z_{Au}\sim20$ \cite{Paul,Gold}.
Therefore, in simulations we assume that the high-Z ions in the microtube are partially photoionized with a mass-to-charge ratio $m_{Au}/Z_{Au}e \simeq 10 m_p/e$, where $m_p$ and $e$ are the mass and charge of proton, respectively.
The simulation box size is $2\lambda \times 2\lambda \times 12\lambda$, the spatial resolution is $\Delta x=\Delta y=\Delta z/2=\lambda/400$, and each cell has 64 macro-particles.
For reference, both laser pulses are assumed to arrive at the microtube center ($x=y=z=0$) at $t=0$, and the simulations begin at $t=-T_0$.

\subsection{Dynamics of a laser-irradiated microtube}
\begin{figure}[htbp]
\includegraphics[width=0.75\textwidth]{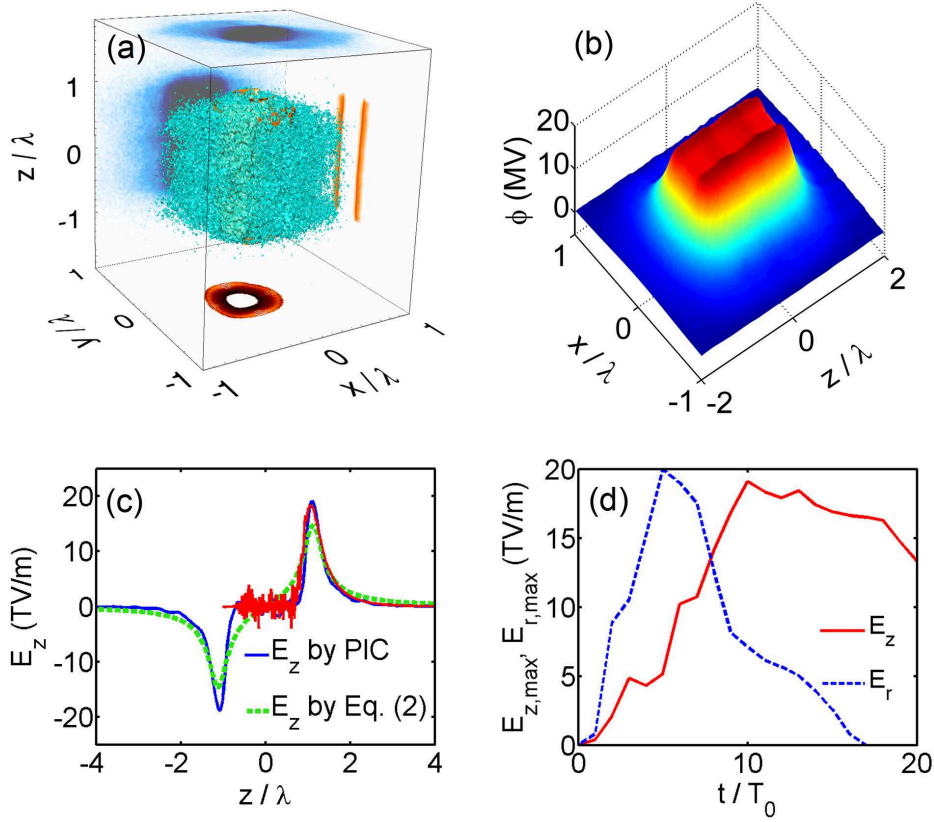}
\caption{ (color online).
(a) Spatial distribution of the electrons (blue) and the ions (orange) at $t=10 T_0$, their density distributions at $z=0$ are respectively shown on the top ($n_e$) and the bottom ($n_i$), $n_e$ at $y=0$ and $n_i$ at $x=0$ are respectively displayed on the rear and right sides of the box.
(b) Three dimensional view of the Coulomb potential $\phi$ at $t=10 T_0$ on the $z=0$ plane.
(c) $E_z$ at $t=10 T_0$ along the $z$-axis obtained from the PIC simulation and the prediction by Eq. (\ref{Ez}), the red solid curve roughly sketches the in-situ field experienced by the injected proton beam that enters the microtube before the arrival of the laser pulse.
(d) Time evolution of the extrema of the axial ($E_{z,\max}$) and the radial ($E_{x,\max}$) electric fields.
} \label{figDynamics}
\end{figure}

To understand the mechanism of laser-irradiated microtube accelerator, we first investigate the dynamics of the electrons and the ions of a laser-irradiated microtube.
Under the irradiation of an intense laser pulse, the electrons of a microtube will be greatly heated due to collisionless absorption mechanisms \cite{ShengCPB}.
Since the thickness of the microtube is on the order of the skin depth $d_{s}\simeq 40\lambda$,
these heated electrons will be largely blown away by the laser radiation \cite{XuAIP2016,XuPOP2017}.
While the ions are left due to their large mass-to-charge ratio as shown in Fig. \ref{figDynamics}(a).
The expanding electrons and the lag-behind ions will conspire to form a saddle-shaped Coulomb potential field in Fig. \ref{figDynamics}(b).
Consequently, quasi-static electric fields are formed both along the longitudinal and radial directions inside the microtube, which can be used for proton acceleration \cite{MurakamiAPL2013} and focusing  \cite{ToncianScience2006}, respectively.
For simplicity, we assume that the laser-irradiated microtube has a uniform net positive charge density $en_0$, and its thickness $D$ is much smaller than its diameter $2R$ and length $2L$. Then the Coulomb potential on the $z$-axis can be written as  \cite{MurakamiAPL2013}
\begin{equation}
\phi(z)=\frac{en_0D R}{2\varepsilon_0} ln\frac{\sqrt{R^2+(z-L)^2}-(z-L)}{\sqrt{R^2+(z+L)^2}-(z+L)}.
\end{equation}
Correspondingly, the axial electric field on the $z$-axis can be solved as
\begin{equation}
E_z=\frac{1}{2\varepsilon_0}[\frac{en_0RD}{\sqrt{(z-L)^2+R^2}}-\frac{en_0RD}{\sqrt{(z+L)^2+R^2}}]. \label{Ez}
\end{equation}
From the simulation, we find that $n_0\simeq 32n_c$ at $t=10 T_0$. Then the above equation predicts that the axial field reaches its extrema $E_{z,\max}=-E_{z,\min}\simeq 15$ TV/m at two orifices of the microtube, which is in good agreement with the simulation result as shown in Fig. \ref{figDynamics}(c).
Inside the microtube, the axial electric field obtained from the simulation is almost zero since the ionized microtube can be considered as a good conductor.
It was proved that this axial electric field can be utilized to accelerate protons initially at rest to MeV energies \cite{MurakamiAPL2013}.

\subsection{Further acceleration of incident proton beams}
More importantly, here we find that there is a response time for the growing of the axial and the radial electric fields. As shown in Fig. \ref{figDynamics}(d), the extrema of the axial and the radial fields reach their peaks at $t\simeq10T_0$ and $t\simeq6T_0$, respectively.
Therefore, if a pre-accelerated proton beam is injected into the microtube along the axis before the rising of the axial field and comes out from another end at around the peak time of the axial field, this proton beam will avoid the decelerating field at the entrance and experience the strong accelerating field at the exit as indicated by the red curve in Fig. \ref{figDynamics}(c). In other words, an injected proton beam can be further accelerated by a laser-irradiated microtube if the time delay between the laser and the proton beam is appropriate.

In simulations, the time delay is defined as the difference in the arrival times of the laser and the proton beams at the microtube center, and it is controlled by the initial position of the proton beam.
Since the axial electric field is very weak inside the microtube, the mean velocity of the proton beam $v_p$ is nearly constant during the proton beam propagation inside the microtube.
Assuming a constant $v_p$ inside the microtube, the time delay between the laser and the proton beam can be calculated as $t_{delay}=T_0+z_0/v_p$, where $z_{0}$ is the z-coordinate of the proton beam center at the initial time $t=-T_0$.
And the times when the proton beam enters (comes out) the microtube can be estimated as $t_{in}=-L/v_p-t_{delay}$ ($t_{out}=L/v_p-t_{delay}$).
Considering the evolution of $E_z$ and $E_r$ displayed in Fig. \ref{figDynamics}(d), we find
\begin{equation}
t_{in}<0,\texttt{ and }t_{out} \sim 10T_0, \label{condition}
\end{equation}
are roughly required for an efficient further acceleration of the injected proton beam. Further, the proton beam should not come out the microtube too late in order to be focused by $E_r$ that peaks at $t\simeq 6T_0$.

\begin{figure}[htbp]
\includegraphics[width=0.75\textwidth]{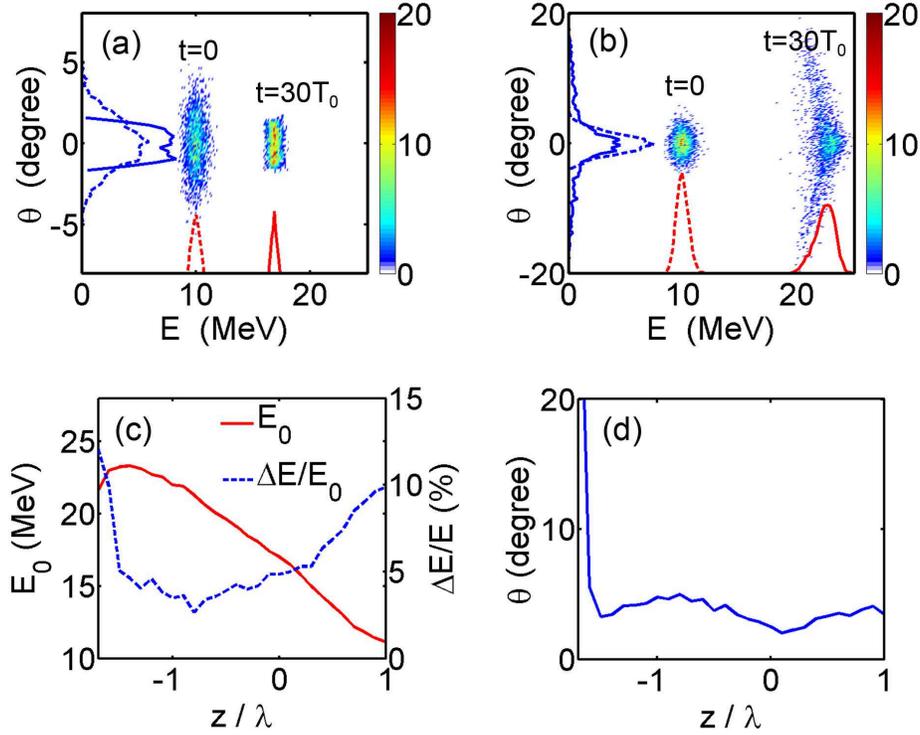}
\caption{ (color online). Comparison the energy-angle distributions (color contours), angular distributions (blue lines) and energy spectra (red lines) of the further accelerated (solid lines) and initially injected (dashed lines) proton beams with different initial positions (a) $z=0\lambda$, (b) $z=-1.6\lambda$.
(c) The mean energy ($E_0$) and energy spread ($\Delta E/E_0$), and (d) the FWHM divergence angle ($\theta$) of the further accelerated proton beam at $T=30T_0$ as functions of the initial z-coordinate $z_0$ of the proton beam.
} \label{figTiming}
\end{figure}

In Fig. \ref{figTiming}, we compare the accelerated proton beams with different initial positions.
In Fig. \ref{figTiming}(a), the proton beam initially centers at $x=y=z=0$ with a size of $40\times40\times40$ nm$^3$ and a density $n_p=n_c$, and its mean energy is $E_0=10$ MeV ($v_{p}\simeq0.145c$ along the axial) with a FWHM spread $\Delta E/E_0=10\%$.
Such an injected proton beam might be obtained from a laser-driven nanotube accelerator \cite{MurakamiAPL2013} or by other laser-driven ion mechanisms with focusing and energy selection \cite{ToncianScience2006}.
Under these parameters, the conditions (\ref{condition}) are satisfied well.
Consequently, the further accelerated proton beam has been enhanced in three aspects.
Firstly, the mean energy increases to $\sim 17$ MeV.
Secondly, the energy spread decreases to $\Delta E/E_0=4.9\%$. Since the accelerating field $E_z$ is at the rising stage when the proton beam comes out the microtube, not only the relative spread $\Delta E/E_0$ but also the absolute spread $\Delta E$ decreases.
Thirdly, the FWHM divergence angle decreases from 3.4 degree to 2.5 degree. This focusing effect can be attributed to the inward radial electric field that peaks at $t\simeq 6T_0$.
In Fig. \ref{figTiming}(b), we set $z_{0}=-1.6\lambda$ ($t_{in}\simeq 3T_0$ and $t_{out} \simeq 16T_0$) and keep other parameters the same as those in Fig. \ref{figTiming}(a).
Then we find that both the energy spread and the divergence of the further accelerated proton beam become worse, although its energy increases.
We show the mean energy and the energy spread, and the FWHM divergence angle of the accelerated proton beam as functions of its initial z-coordinate $z_0$ in Fig. \ref{figTiming}(c) and (d), respectively.
It illustrates that the obtained energy increases with the decreasing $z_0$ in the region $-L<z_0<L$, since a smaller $z_0$ is equivalent to a stronger accelerating field at the exit in this region.
However, if $z_0$ is too small, the quality of the accelerated proton beam will quickly deteriorate since now it experiences a strong decelerating field at the entrance of the microtube. In addition, the radial electric field also becomes too weak to focus the proton beam if $z_0$ is too small.
From a series of simulations, we find the injected proton beam can be efficiently further accelerated and maintain its quality with $-L\leq z_0\leq 0$.

\section{Discussion and Conclusion}
\begin{figure}[htbp]
\includegraphics[width=0.75\textwidth]{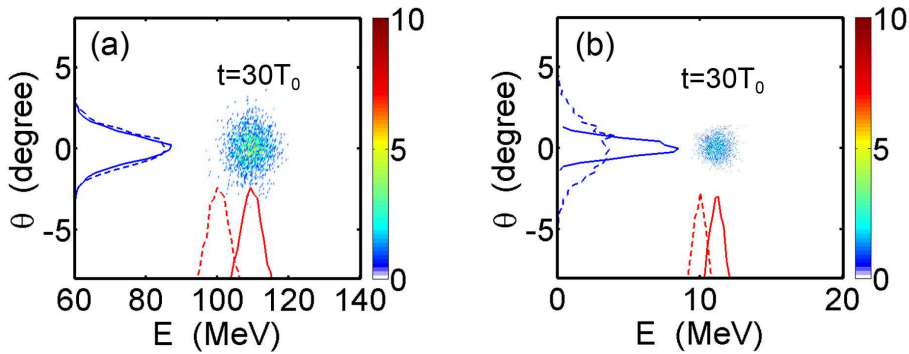}
\caption{ (color online).
Energy-angle distributions (color contours), angular distributions (blue lines) and energy spectra (red lines) of the further accelerated (solid lines) and initial (dashed lines) proton beams in the cases with (a) a higher initial mean energy $E_0=100$ MeV, (b) a moderate laser intensity $a_0=0.5$. The other parameters are similar to those in Fig. \ref{figTiming}(a).} \label{figDiscuss}
\end{figure}

In principle, this acceleration scheme can be applied to further accelerate the injected proton beam with initial energies up to a few hundred MeV, and the high-quality further accelerated proton beam can be employed as the injected beam for the next acceleration.  Therefore, this acceleration scheme is particularly suitable for multi-stage acceleration of protons.
In Fig. \ref{figDiscuss}(a), the further acceleration of a more energetic proton beam ($E_0=100$ MeV) is evidenced. In this simulation, the proton beam initially centers at $z_0=-L$ and the microtube is lengthened up to $2L=4\lambda$ to guarantee the condition (\ref{condition}) for the more energetic protons. Consequently, the mean energy of protons increases to 109 MeV, and the high-quality of the proton beam is well maintained.
We notice that the laser-to-proton energy conversion efficiency is very low ($\sim 10^{-5}$) in the above simulations. Firstly, this problem may be partially alleviated by use of an array of microtubes in the laser focal region instead of an isolated microtube \cite{Andreev}.
Secondly, we find surprisingly that the energy conversion efficiency can be greatly enhanced by using a non-relativistic laser pulse.
In Fig. \ref{figDiscuss}(b), we set a moderate laser intensity $a_0=0.5$ and other parameters the same as those in Fig. \ref{figTiming}(a). In this case the laser energy ($\propto a_0^2$) is reduced by two order of magnitude, while the increase in the proton mean energy is reduced by only one order of magnitude to about 1 MeV as shown in Fig. \ref{figDiscuss}(b).
As a result, the laser-to-proton energy conversion efficiency is increased to $\sim 10^{-4}$.
Further the laser pulses at non-relativistic intensities could have higher repetition rates and be much more available and economic than the relativistic ones.
Thirdly, the energy conversion efficiency might be also improved by using a larger microtube, which allows more initially injected protons. Note that the size of the microtube in our simulations is close to the limit of our computational capacity. The simulations using larger microtubes would require extremely large computational resources and should be further studied in the future.

In summary, a cascaded ion acceleration scheme using laser-irradiated microtubes is proposed and verified by numerical simulations. By controlling the arrival time of an injected proton beam, this proton beam can be efficiently further accelerated by the transient axial electric field in the laser-irradiated microtube. Further, the energy spread of the injected proton beam can be reduced if the accelerating field is still rising when the proton beam comes out the microtube. In addition, the divergence of the injected proton beam could be suppressed by the inward radial electric field in the microtube.
More importantly, this cascaded acceleration scheme works well with the non-relativistic laser pulses and the injected proton beams with initial energies up to a few hundred MeV.
Therefore, proton beams may be accelerated in multi-stages with this scheme to higher energy.

\begin{acknowledgments}
This work was supported in part by the National Basic Research Program of China (Grant No. 2013CBA01504) and the National Natural Science Foundation of China (Grant Nos. 11675108, 11421064, 11405108 and 11374210). S.M.W. and M.C. appreciate the supports from National 1000 Youth Talent Project of China. Z.M.S acknowledges the support of a Leverhulme Trust Research Project Grant at the University of Strathclyde. Simulations have been carried out at the PI cluster of Shanghai Jiao Tong University.
\end{acknowledgments}


\bibliography{apssamp}

\end{document}